# Can Other People Make You Less Creative?

Liane Gabora

Department of Psychology, University of British Columbia
Okanagan Campus, Arts Building, 333 University Way, Kelowna BC, V1V 1V7, CANADA

## Abstract

This paper explains in layperson's terms how an agent-based model was used to investigate the hypothesis that culture evolves more effectively when novelty-generating creative processes are tempered by imitation processes that preserve proven successful ideas. Using EVOC, an agent-based model of cultural evolution we found that (1) the optimal ratio of inventing to imitating ranged from 1:1 to 2:1 depending on the fitness function, (2) there was a trade-off between the proportion of creators to conformers and how creative the creators were, and (3) when agents in increased or decreased their creativity depending on the success of their latest creative efforts, they segregated into creators and conformers, and the mean fitness of ideas across the society was higher. It is tentatively suggested that through the unconscious use of social cues, members of a society self-organizes to achieve a balanced mix of creators and conformers.

Do you feel more creative in some social environments than others, or even when you are completely alone? Although it is widely believed that stimulating environments enhance creativity there is evidence that this is not always the case; indeed it turns out that creativity can be enhanced by spending time in an isolation tank blocked off from sensory stimulation entirely (Forgays & Forgays, 1992; Norlander, Bergman, & Archer, 1998; Vartanian, & Suedfeld, 2011). This suggests that social stimulation may not always enhance creativity, and may even decrease it.

The idea that social environments could interfere with creativity came from thinking about creativity in the context of culture as an evolutionary process. Evolutionary processes require a balance between exploratory processes that generate new variants and conservative processes that perpetuate 'tried and true' variants. This is the case in biological evolution, and it seemed reasonable that it would also be the case for cultural evolution. In other words: people may send out social signals to each other that ensure that creativity---the process that fuels cultural novelty---is properly balanced by conformity, i.e., imitation of ideas that have already proven to be successful.

The notion that society might aim to temper novelty-generating creativity with novelty-perpetuating imitation flies in the face of the widespread assumption that more creativity is necessarily better. It goes without saying that creativity is a good thing, and that everyone should be creative. Or does it? Sure, our capacity for self-expression, for finding practical solutions to problems of survival, and coming up with aesthetically pleasing objects that delight the senses, all stem from the creative power of the human mind. But there are drawbacks to creativity.

First, creative people tend to be more emotionally unstable and prone to affective disorders such as depression and bipolar disorder. They have a higher incidence of 'schizotypal' leanings than other segments of the population. They are also more prone to abuse drugs and alcohol, and to commit suicide. So there is a 'dark side' to creativity (Cropley, Cropley, Kaufman, & Runco, 2010).

Second, a creative solution to one problem often generates other problems, or unexpected negative side effects that may only become apparent after much has been invested in the creative solution. There is a cultural version of what in biology is referred to as epistasis, where what is optimal with respect to one part depends on what is done with respect to another part. Once both parts of a problem have been solved in a mutually beneficial way, too much creativity can cause these 'co-adapted' partial solutions to break down.

Third, in a group of interacting individuals, only a fraction of them need be creative for the benefits of creativity to be felt throughout the group. Uncreative people can reap the benefits of the ideas of 'creative types' without having to withstand the 'dark side' of creativity by simply imitating, or admiring them. Few of us know how to build a computer, or write a symphony, or a novel, but they are nonetheless ours to use and enjoy when we please. An excess of creative types all completely absorbed in their own creative process might effectively insulate themselves and block the rapid diffusion of the best ideas.

This opens up some interesting questions. Would it be good for the society as a whole if everyone were highly creative? In order for a culture to evolve optimally, what is the ideal ratio of creators to imitators, and how creative should the 'creative types' be? And perhaps most interesting of all: do people upgrade or downgrade how creative they are in response to social cues they receive from other people about the perceived value of their creative outputs?

My colleagues and I are investigating these questions using a computer model of cultural evolution. I'll begin by telling you a bit about the computer model itself. Then I'll explain the experiments that led up to the experiments that explored the hypothesis that people socially regulate each others' creativity. Finally, I'll explain the social regulation experiments themselves.

THE COMPUTER MODEL

The current model's predecessor was called Meme and Variations or MAV (Gabora, 1995). Its name is a pun on the musical form, 'theme and variations'. MAV was the earliest computer program to model culture as an evolutionary process in its own right. MAV was inspired by the genetic algorithm (GA), a search technique that finds solutions to complex problems by generating a 'population' of candidate solutions (through processes akin to mutation and recombination), selecting the best, and repeating until a satisfactory solution is found.

The computer model is composed of an artificial society of agents in a two-dimensional grid-cell world. Agents consist of (1) a neural network, which encodes ideas for actions and detects trends in what constitutes an effective action, and (2) a body, which implements their ideas as actions. The agents can do two things: (1) invent ideas for new actions, and (2) imitate their neighbors' actions. The computer model enables us to investigate what happens to the diversity and effectiveness of actions in the artificial society over successive rounds (called 'iterations') of invention and imitation. Since the ideas in the model are ideas for actions, diversity is measured by counting how many different actions are being implemented by the agents. Evolution in the biological sense is not taking place; the agents neither die nor have offspring. But evolution in the cultural sense is taking place through the generating and sharing of ideas for actions amongst agents, which over time leads to more effective actions.

In MAV, all agents were equally capable of both inventing and imitating. In the latest version of the computer model called EVOC (for EVOlution of Culture), it is possible to vary how likely an agent is to invent versus imitate.

A TYPICAL RUN

Each iteration, every agent has the opportunity to (1) acquire an idea for a new action, either by imitation, copying a neighbor, or by invention, creating one anew, (2) update their knowledge about what constitutes an effective action, and (3) implement a new action. Effectiveness of actions starts out low because initially all agents are just standing still doing nothing. Soon some agent invents an action that has a higher effectiveness than doing nothing, and this action gets imitated, so effectiveness increases. Effectiveness increases further as other ideas get invented, assessed, implemented as actions, and spread through imitation. The diversity of actions initially increases due to the proliferation of new ideas, and then decreases as agents hone in on the fittest actions. Thus MAV successfully models how 'descent with modification' can occur in a cultural context.

FIRST SET OF EXPERIMENTS:
WHAT IS THE OPTIMAL RATIO OF EFFORT INVESTED IN CREATIVITY TO EFFORT INVESTED IN IMITATION?

In the earliest version of this computer model (MAV), all agents were equally capable of both inventing and imitating (Gabora, 1995). It was possible to vary the probability that, in a given iteration, they would invent versus imitate. The agents are too rudimentary to suffer from depression or commit suicide, so it is just the other detrimental aspects of creativity listed above that we thought might play a role in these experiments.

What the results showed was that if they all imitated each other all the time, nothing happened at all: everyone watched everyone else and no one did anything. If they all invented all the time, the progress of ideas was slow because they weren't taking advantage of each other's hard work. The optimal ratio of inventing to imitating was about 2:1. Of course, these results hold for just this little idealized world, and they may or may not be generalizable to the world at large. But they do ring true for many people, who say they spend about 2/3 of their time alone in their studio or office immersed in their work, and about 1/3 of their time talking or reading about or studying things related to their creative project. (Subsequent experiments showed that this ratio can be as low as 1:1 depending on the 'fitness function', i.e., the kind of task the agents had to solve.)

SECOND SET OF EXPERIMENTS:

WHAT PROPORTION OF SOCIETY SHOULD BE CREATIVE, AND HOW CREATIVE SHOULD THEY BE?

The finding that very high levels of creativity can be detrimental for society led to the hypothesis that there is an adaptive value to society's ambivalent attitude toward creativity; society as a whole may benefit from a distinction between the conventional workforce and what has been called a "creative class" (Florida, 2002). Using the new version of the computer model, EVOC we investigated how patterns of cultural evolution are affected by how numerous and creative the creators are. We conducted experiment that varied, not just the ratio of creators to imitators, but also the creativeness of the creators (Leijnen & Gabora, 2009). Each agent could be a pure imitator, a pure creator, or something in between. The pure imitators never invented; they simply copied the successful innovations of the creative agents. The creators were able to invent as well as well as imitate. The percentage of iterations in which they invented varied up to 100%.

The results were provocative. We found that cultural diversity – that is, the number of different actions in the artificial society -- was positively correlated with both the percentage of creators, and their level of creativity. However, for cultural fitness or effectiveness, the situation was more complex. So long as the creative types weren't THAT creative, the more of them there are, the better. But when the creative types were HIGHLY creative, the mean fitness of ideas across the society was higher if there were fewer of them. The results seemed to show that the more creative the creators are, the less numerous they should be.

We then conducted more extensive investigations of these questions employing more detailed and sophisticated analytical methods. The amount of time it takes for the effectiveness of ideas across the artificial society to reach a threshold level of performance is affected by the creator to imitator ratio (C), and the creator innovation probability (p).

The same general trends emerged. Cultural diversity was once again positively correlated with both the percentage of creators, and their level of creativity (Gabora & Leijnen, 2013). But when we looked at, not the variety of ideas, but how fit or effective they were, the pattern of results was more complicated. What we found was that if C is low the p should be high, but if C is high then p should be intermediate (Gabora & Firouzi, 2012). In other words, once again there was a tradeoff between how creative the creators were, and how many of them there should be.

These results supported the hypothesis that too much creativity causes 'co-adapted' partial solutions to problems to break down (the 'don't fix it if it ain't broke' phenomenon). They also supported the hypothesis that creative types, while they are a necessary source of novelty, constitute pinholes in the fabric of culture that block the spread of ideas. An iteration spent inventing is an iteration not spent imitating, and imitation is extremely valuable. It's not just a form of free-riding, nor just 'the greatest compliment', but an indispensable social mechanism that serves everyone. By simply copying the successful innovations of the creative types, imitators serve as a 'memory' for preserving the fittest configurations. So, contrary to popular belief, it might not be best for the society as a whole if everyone were creative.

THIRD SET OF EXPERIMENTS:
HOW IS SOCIETY AFFECTED IF PEOPLE CAN UPGRADE/DOWNGRADE THEIR

CREATIVITY IN RESPONSE TO SOCIAL SIGNALS?

We then hypothesized that society as a whole might perform even better with the ability to adjust creativity in accordance with their perceived creative success, through mechanisms such as selective ostracization of deviant behaviour unless accompanied by the generation of valuable cultural novelty, and encouraging of successful creators. A first step in investigating this hypothesis was to determine whether it is algorithmically possible to increase the mean fitness of ideas in a society by enabling them to self-regulate how creative they are. To test the hypothesis that the mean fitness of cultural outputs across society increases faster with social regulation (SR) than without it, we increased the relative frequency of invention for agents that generated superior ideas, and decreased it for agents that generated inferior ideas (Gabora & Tseng, 2014a). Each iteration, for each agent, the fitness of its current action relative to the mean fitness of actions for all agents at the previous iteration was assessed. If its action was fitter than the mean it created more, and if its action was less fit than the mean it imitated more.

The typical pattern was observed with respect to the diversity, or number of different ideas: an increase as the space of possibilities is explored followed by a decrease as agents converge on fit actions. However, this pattern occurred earlier, and was more pronounced, in societies with SR than societies without it. Interestingly, the mean fitness of the cultural outputs in societies with SR was higher than that in societies without SR.

Even more interestingly, the societies with SR ended up separating into two distinct groups: one that primarily invented, and one that primarily imitated. Thus the observed increase in fitness could indeed be attributed to increasingly pronounced individual differences in their degree of creative expression over the course of a run. Agents that generated superior cultural outputs had more opportunity to do so, while agents that generated inferior cultural outputs became more likely to propagate proven effective ideas rather than reinvent the wheel.

We conducted another set of experiments in which agents were able to generate not just simple single-step actions but complex multi-step actions that were more like real-world actions such as dancing or tool-making (Gabora & Tseng, 2014b). The same general patterns emerged: the agents segregated into creators and conformers, and the mean fitness of actions across the society as a whole increased.

These results don't show that this kind of segregation into creators and conformers actually takes place in real societies. What they suggest is that if this did happen it could have an adaptive benefit for society as a whole.

CONCLUSIONS

The computer model differs substantially from the real world, so any attempt to extrapolate these results should be taken with a grain of salt. The agents' neural networks are so small that creative novelty is generated does not involve noticing and refining new kinds of connections the way it happens in real minds (Gabora, 2000). Moreover, in these simulations, unlike the real world, agents had only one task to accomplish. However, the results of these computer simulations are provocative, and inspire new ways of thinking about creativity. They show that societies may benefit as a whole by self-organizing into a balanced mix of novelty generating creators and continuity perpetuating imitators, and lead to the speculation that this happens spontaneously in real societies.

There are many speculative but fascinating implications of this research. It leads to the suggestion that the reason creative individuals often isolate themselves is not just to decrease disturbances so they can more fully concentrate on their art, but because isolation safeguards you from social signals to downgrade your creativity, which can negatively impact creative performance. It suggests that as societies become increasingly denser it may become increasingly difficult to isolate oneself from the cues by which people unconsciously socially regulate each other's level of creativity.

Another possibility, suggested by my colleague Kiley Hamlin, is that girls are more responsive to social cues to downgrade creativity than boys, which might prematurely streamline them into an "imitator" track. The more you imitate, the more you rely on imitation as a source of ideas and ways of doing things, so it becomes a vicious circle of sorts, and vice versa: the more you create, the more you keep the creative juices flowing and rely on your own creative processes as a source of not just new ideas, but pleasure.

ACKNOWLEDGMENTS

The author is grateful for funding from the Natural Sciences and Engineering Research Council of Canada.